**Title**

# Controlling placement of nonspherical (boomerang) colloids in nematic cells with photopatterned director

**Authors**

Chenhui Peng[1], Taras Turiv[1], Rui Zhang[2], Yubing Guo[1], Sergij V. Shiyanovskii[1], Qi-Huo Wei[1], Juan de Pablo[2] and Oleg D. Lavrentovich[1]*

Corresponding Author: *olavrent@kent.edu

**Affiliations**

[1]Liquid Crystal Institute and Chemical Physics Interdisciplinary Program, Kent State University, Kent, OH 44242, USA

[2]Institute for Molecular Engineering, University of Chicago, Chicago, Illinois 60637, USA

**Abstract**

Placing colloidal particles in predesigned sites represents a major challenge of the current state-of-the-art colloidal science. Nematic liquid crystals with spatially varying director patterns represent a promising approach to achieve a well-controlled placement of colloidal particles thanks to the elastic forces between the particles and the surrounding landscape of molecular orientation. Here we demonstrate how the spatially varying director field can be used to control placement of non-spherical particles of boomerang shape. The boomerang colloids create director distortions of a dipolar symmetry. When a boomerang particle is placed in a periodic splay-bend director pattern, it migrates towards the region of a maximum bend. The behavior is contrasted to that one of spherical particles with normal surface anchoring, which also produce dipolar director distortions, but prefer to compartmentalize into the regions with a maximum splay. The splay-bend periodic landscape thus allows one to spatially separate these two types of particles. By exploring overdamped dynamics of the colloids, we determine elastic driving forces responsible for the preferential placement. Control of colloidal locations through patterned molecular orientation can be explored for future applications in microfluidic, lab on a chip, sensing and sorting devices.

**MAIN TEXT**

**Introduction**

Hybrid soft matter systems, representing colloidal particles dispersed in a nematic liquid crystals (LCs) (*1-7*) have gained a significant interest over the last decades. The reason is that orientational order of the nematic medium and anisotropy of surface energy at the colloid-nematic interfaces makes the colloidal interactions long-range and anisotropic, thus producing a broad spectrum of ordered patterns. The LC orientational order leads to long-range anisotropic colloidal interactions which can be used to form linear chains, anisotropic clusters, 2D and 3D periodic colloidal assemblies through interactions with flat, curved, templated interfaces and topological defects (*3, 5, 7-19*). The LC mediated colloidal self-assembly further expands the spectrum of possible design pathways when combined with the external stimuli such as electromagnetic fields, temperature, or confinement.

. Usually, the patterns of colloidal assemblies in LCs are considered for two extreme cases, with the background representing (i) a uniform director field (*1-4, 6, 7*) or (ii) containing topological defects that attract colloidal inclusions to their cores (*5, 11-13, 15, 20-25*). The intermediate case, when the typical scale of director gradients is larger than the size of the colloid, was explored only very recently (*26-29*). Experimental studies of this intermediate case have been enabled by the development of a robust patterned photoalignment technique (*28*). In this technique, one irradiates a photosensitive substrate through a plasmonic photomask with a predesign pattern of submicrometer slits; the passing light becomes locally polarized. This polarization pattern is being imprinted into the photosensitive substrate that is used to align the liquid crystal. The approach allows one to create any pattern of the director field with the typical scale of spatial gradients ranging from about 1 micrometer to hundreds of micrometers (*28*). As demonstrated by Peng et al(*29*), photopatterned nematic cells with the period larger than the colloidal size can be used to guide the spatial placement of spherical colloids with different surface anchoring. The goal of this work is to extend the study to the case of non-spherical particles, such as boomerang-shaped colloids. Controlled assembly of nonspherical colloids is a subject of current studies for both isotropic (*30*) and LC environments (*31, 32*). One of the challenging tasks is to force the particles to occupy certain spatial locations and to align themselves along a preferred direction. In this work, we demonstrate that both the location and alignment of the non-spherical particles such as boomerang colloids can be achieved by using spatially varying director fields, such as splay-bend periodic patterns, Moreover, the predesigned patterns of molecular orientations can also be used to separate particles of different shape, even if the director distortions created by them are of the same symmetry (dipolar in our case).

**Results**

We use nematic LCs, in which elongated molecules are oriented along a single direction called the director $\hat{\mathbf{n}}$, with the properties $\hat{\mathbf{n}} \equiv -\hat{\mathbf{n}}$ and $\hat{\mathbf{n}}^2 = 1$. The flat nematic cells bounded by two glass substrates which have identical director field of periodic one-dimensional splay $\text{div}\hat{\mathbf{n}} \neq 0$ and bend $\hat{\mathbf{n}} \times \text{curl}\hat{\mathbf{n}} \neq 0$, modulated along the $y$-axis, Fig.1a. The director field imposed by the surface photo-alignment contains only splay and bend; there is no twist nor saddle-splay:

$$\hat{\mathbf{n}} = (n_x, n_y, n_z) = (\cos\alpha, \sin\alpha, 0), \qquad (1)$$

where $\alpha = \pi y/l$, $l = 80\,\mu\text{m}$ is the period, Fig. 1. There is no $z$-component, as the director is everywhere parallel to the bounding plates. The experimental cell represents a thin (thickness $h = 20\,\mu\text{m}$) flat layer of a nematic sandwiched between two glass plates with identical photo-induced alignment patterns at the top ($z = h$) and the bottom ($z = 0$) surfaces. Figure 1a shows the spatial distribution of the dimensionless bend energy density $(l/\pi)^2 (\hat{\mathbf{n}} \times \text{curl}\hat{\mathbf{n}})^2 = \sin^2(\pi y/l)$. As detailed by the experiments below, the alternating regions of splay and bend attract colloidal particles with different shapes, namely, boomerang colloids with tangential surface anchoring are attracted to the bend regions, Figs. 1-4, while spheres with perpendicular anchoring are attracted to the splay regions, Fig. 5.

**Placement of boomerang colloids**. The boomerang-shaped colloids are fabricated from the epoxy-based photoresist SU-8 following the procedure in Ref.(*33*), with symmetric arms of length $a = 2.1\,\mu m$ each, thickness $0.51\,\mu m$, width $0.55\,\mu m$, and an apex angle $100 \sim 110°$. The LC director aligns parallel to the surface of the particles; we call such an alignment tangentially degenerate, or simply tangential. Individual boomerang colloids placed in the cell with a periodic splay and bend migrate towards the regions with the maximum bend, $y = \pm l/2, \pm 3l/2,...$, Fig. 1a. When the concentration of boomerangs is sufficiently high, they form curved chains extended along the $y$-axis, Fig. 1b. One might expect that the boomerang particles, creating director distortions of a dipolar type, would also form chains extended along the $x$-axis. One of the reasons that the clusters extend along the $y$-axis might be the kinetics of their mutual approach, which is driven by the elastic forces acting along the $y$-axis; furthermore, the opening angle of the used boomerang particles is relatively large which weakens the dipolar interactions. Smaller opening angles might favor assemblies along the $x$-axis. In the future, it will be of interest to explore whether the variations of the opening angle can be used to control the scenario of assemblies.

Below, we characterize the behavior of individual boomerang-shaped particles in the distorted field of the splay-bend stripes. If one places a boomerang colloid by optical tweezers in any location within the pattern specified by Eq. (1) and then release it, the boomerang moves towards the regions $y = \pm l/2, \pm 3l/2,...$, while simultaneously rotating the arms in accordance with the rotating local director, Fig. 1c. This dynamics can be explored in order to deduce the details of the interaction potential between the particle and the underlying landscape of director deformations.

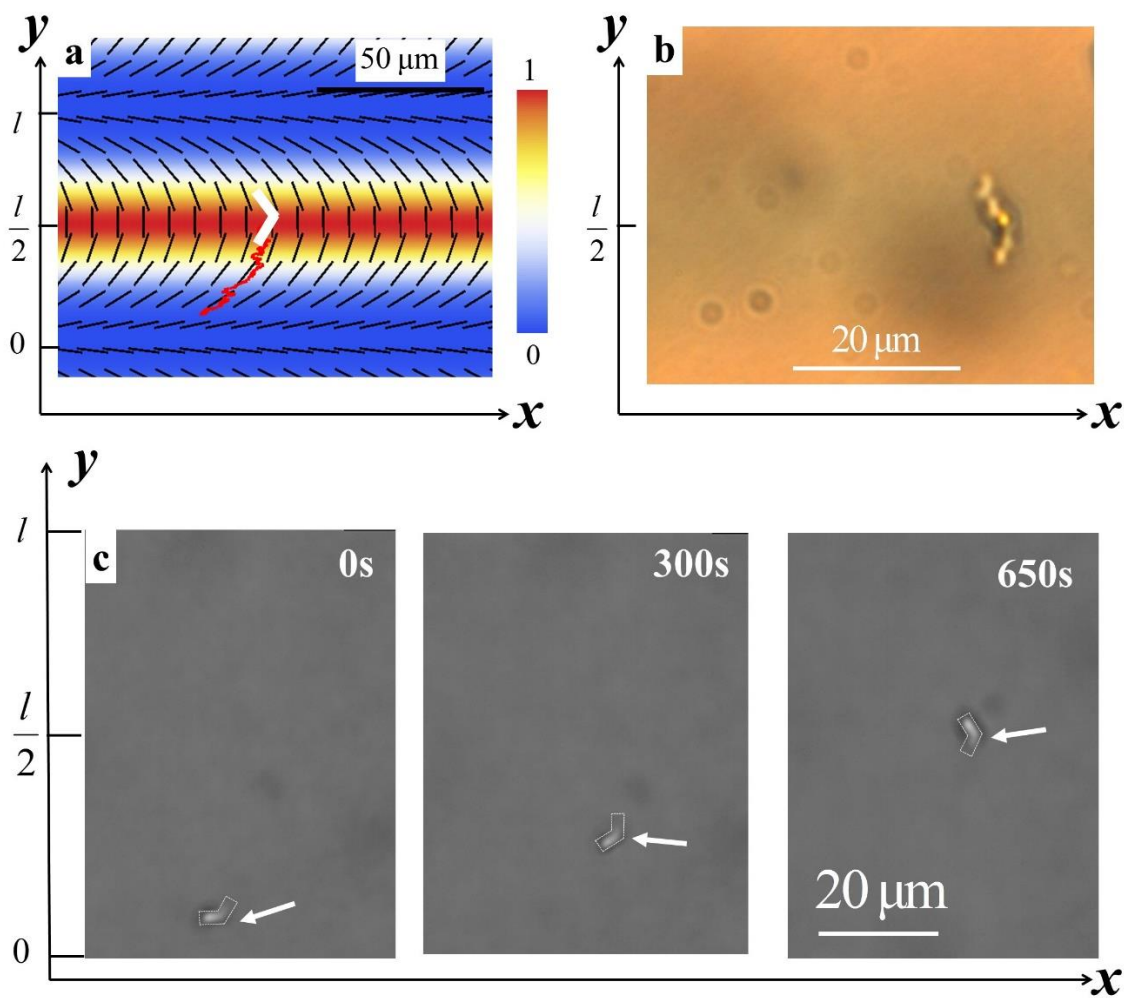

**Fig.1. Elasticity-directed placement and assembly of boomerang colloids in regions with bend deformations of the patterned director.** (**a**) Periodic splay-bend stripe pattern; the normalized bend energy density is labeled by colors according to the scale on the right hand side. A boomerang colloid placed in the splay region by optical tweezers and released there, migrates towards the bend region. The dark red curve is the typical experimental trajectory of the colloid's center. (**b**) Self-assembly of colloids into linear chains in the regions of maximum bend (bright field microscope, unpolarized light); (**c**) Microscopy images of a boomerang colloid moving from the splay region to the bend region, with concomitant reorientation of the colloidal arms; the white arrow indicates the

apex of the boomerang. The period of patterned director is $l = 80\ \mu m$. The LC material used is MLC6815.

The interaction potential between a boomerang colloid and a splay-bend director for our geometry follows from the model proposed by Pergamenshchik (*26*):

$$U_\gamma = -8\pi\gamma K a^2 \sin\alpha \frac{d\alpha}{dy}, \qquad (3)$$

where $a$ is the arm length of the boomerang colloid, $\gamma > 0$ is an unknown numerical coefficient that depends on the shape of the boomerang (in particular, its opening angle) as well as on the anchoring strength of its surface, $K$ is the elastic constant in the so-called one-constant approximation. For the director field in Eq. (1), $\alpha = \pi y / l$, and with $y$ changing in the range from 0 to $l/2$, the potential writes $U_\gamma = -8\pi^2 \gamma K \frac{a^2}{l} \sin\frac{\pi y}{l}$. The last expression for the potential predicts that the boomerang particle should migrate from the region of maximum splay, $y = 0$, towards the locations of maximum bend, $y = l/2$, as observed in the experiments, Fig. 1c.

The force driving the boomerang colloids to the equilibrium location is

$$F_\gamma = -\left(\partial U_\gamma / \partial y\right) = 8\pi^3 \gamma K \frac{a^2}{l^2} \cos\frac{\pi y}{l}. \qquad (4)$$

Since the Reynolds number in our experiments is small, the driving force is balanced by the viscous drag force $\mathbf{F}_\gamma = -\overline{\zeta}\mathbf{v}$, or

$$\mathbf{F}_\gamma = \zeta_\parallel v_\parallel \left(\hat{\mathbf{e}}_x \cos\alpha + \hat{\mathbf{e}}_y \sin\alpha\right) + \zeta_\perp v_\perp \left(-\hat{\mathbf{e}}_x \sin\alpha + \hat{\mathbf{e}}_y \cos\alpha\right), \qquad (5)$$

where $\overline{\zeta}$ is the diagonal friction tensor with the components $\zeta_\parallel$ and $\zeta_\perp$, $\mathbf{v}$ is the particle velocity with the components $v_\parallel$ and $v_\perp$; the subscripts $\parallel$ and $\perp$ refer to the components parallel and perpendicular to the director, respectively.

By balancing the driving and drag forces in the laboratory coordinate system, we obtain the equations for the velocity components $v_x = v_\parallel \cos\frac{\pi y}{l} - v_\perp \sin\frac{\pi y}{l}$ and $v_y = v_\parallel \sin\frac{\pi y}{l} + v_\perp \cos\frac{\pi y}{l}$:

$$\frac{\partial x}{\partial t} = \left(-8\pi^3 \gamma K \frac{a^2}{l^2} \sin\frac{\pi y}{l} \cos^2\frac{\pi y}{l}\right)\left(\frac{1}{\zeta_\parallel} - \frac{1}{\zeta_\perp}\right), \tag{6}$$

$$\frac{\partial y}{\partial t} = \left(-8\pi^3 \gamma K \frac{a^2}{l^2} \cos\frac{\pi y}{l}\right)\left(\frac{\sin^2\frac{\pi y}{l}}{\zeta_\parallel} + \frac{\cos^2\frac{\pi y}{l}}{\zeta_\perp}\right). \tag{7}$$

Solutions of Eq. (6) and Eq. (7) establish how the boomerang's coordinates $x$ and $y$ vary with time. These theoretical dependencies are used to fit the experimental trajectory in Fig. 1a in order to verify the model and to deduce the unknown parameter $\gamma$. Note that the theory is based on the assumption that the particle has enough time to reach an equilibrium orientation at any transient location; Fig.1c illustrates clearly such an adjustment of the particle to the local director field. This assumption is justified by the fact that the rotational motion is much faster than the translational motion, since the characteristic length of patterned director distortions is much larger than the size of the particle.

Fitting of the experimental dynamics of the boomerang colloids requires a knowledge of the two friction components, $\zeta_\parallel$ and $\zeta_\perp$. These two were determined experimentally for the same LC and boomerang colloids by measuring the mean square displacements (MSD) as a function of time for the Brownian diffusion parallel and perpendicular to the director in uniformly aligned cells, $\hat{\mathbf{n}}_0 = (1,0,0)$, Fig. 2. The slopes of the MSD vs. time dependencies yield the diffusion coefficient $D_\parallel = (1.71 \pm 0.1) \times 10^{-14}$ m²/s for the displacements parallel to $\hat{\mathbf{n}}_0$ and $D_\perp = (0.21 \pm 0.02) \times 10^{-14}$ m²/s for the perpendicular displacements. These values yield $\zeta_\parallel = k_B T / D_\parallel = (2.46 \pm 0.05) \times 10^{-7}$ N·s/m and $\zeta_\perp = k_B T / D_\perp = (1.9 \pm 0.1) \times 10^{-6}$ N·s/m.

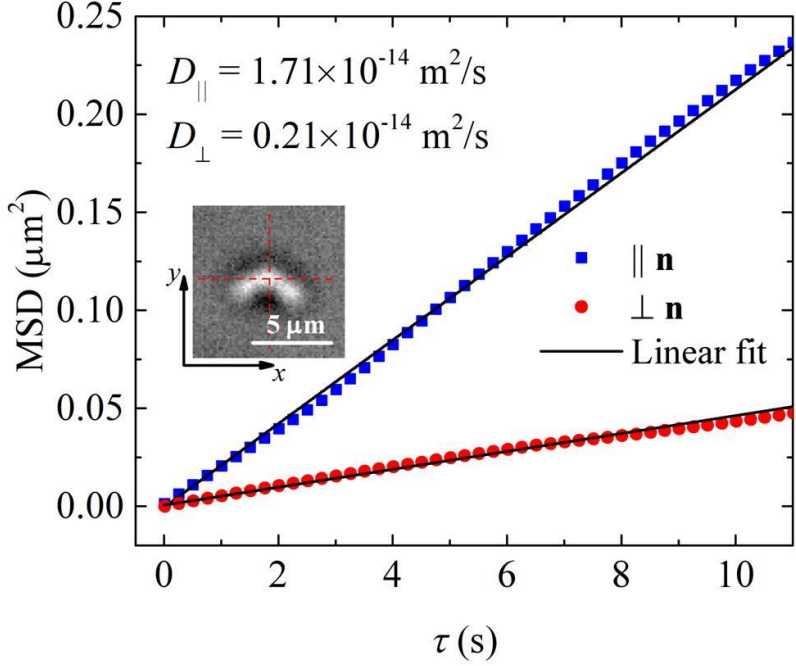

**Fig. 2. Mean squred displacement (MSD) of a boomerang colloid in uniformly aligned LC cell**. Boomerang colloid in a LC (MLC6815) cell with uniform alignment $\hat{\mathbf{n}}_0$, which is achieved by rubbing the aligning agent of PI2555 on the surface.

As clearly seen from Fig. 3a and b, the experimental data are closely matched by the expected theoretical behavior. The only fitting parameter is $\gamma = 0.018 \pm 0.001$. With this value and with $K = 10^{-11}$ N, the depth of the potential minimum, $U_{\gamma,\max} - U_{\gamma,\min} = 8\pi^2 \gamma K \left( a^2 / l \right)$ is $7.8 \times 10^{-19}$ J, i.e., about two orders of magnitude higher than the thermal energy $k_B T = 4.2 \times 10^{-21}$ J. One can also estimate the stiffness of the potential trap that keeps the boomerang colloids in the bend regions. For $y \to l/2$, the stiffness of the elastic trap $k = \dfrac{\partial^2 U_d}{\partial y^2} = 8\pi^4 \gamma \dfrac{a^2}{l^3} K$ is about $1.2 \times 10^{-9}$ N/m.

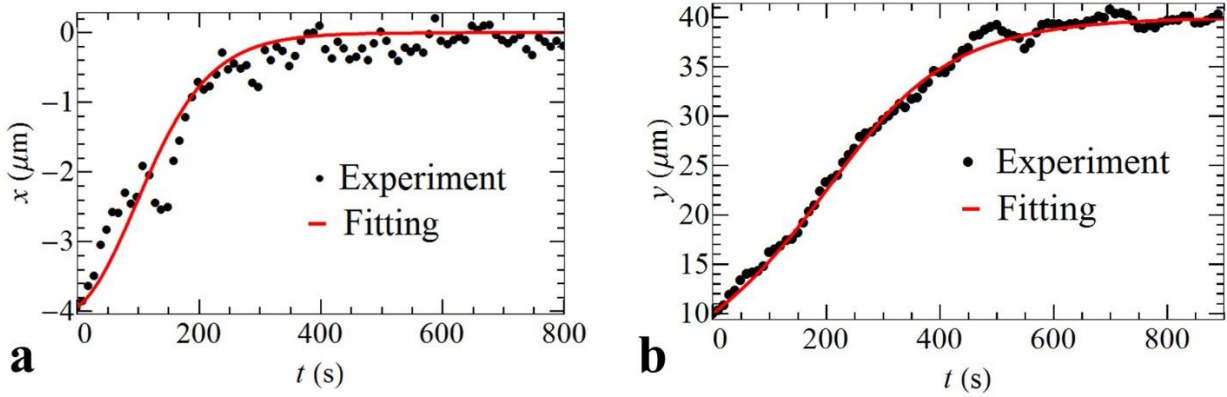

**Fig. 3. Analysis of dynamics of boomerang colloid migrating from splay to bend region**. (**a**) Experimentally measured $x(t)$ dependence for a boomerang colloid moving from $y=0$ to $y=l/2$ and its theoretical fit by Eq. (6). (**b**) Experimentally measured $y(t)$ dependence for a boomerang colloid moving from $y=0$ to $y=l/2$ and its theoretical fit by Eq. (7). LC used is MLC6815.

**Simulations of a boomerang colloid in a photo-patterned cell.** It is of interest to supplement the analytical consideration above with the numerical simulations in which factors such as finite surface anchoring can be included explicitly. To simulate the boomerang with finite surface anchoring, we introduce a tensorial order parameter **Q** for the nematic LC, defined as $\mathbf{Q} = \langle \mathbf{nn} - \mathbf{I}/3 \rangle$, with **I** the identity matrix and $\langle ... \rangle$ the ensemble average. The system is then described by the Landau-de Gennes free-energy model. The free energy reads (*34*)

$$E = \int_{bulk} \left\{ \frac{A}{2}\left(1 - \frac{U}{3}\right) Q_{ij}Q_{ij} - \frac{AU}{3} Q_{ij}Q_{jk}Q_{ki} + \frac{AU}{3}\left(Q_{ij}Q_{ij}\right)^2 \right\} dV + \int_{bulk} \frac{L}{2}\frac{\partial Q_{ij}}{\partial x_k}\frac{\partial Q_{ij}}{\partial x_k} dV + \int_{surface} \left\{ W\left(\tilde{Q}_{ij} - \tilde{Q}_{ij}^{\perp}\right)^2 \right\} dS. \quad (8)$$

where $A$ is the scale of energy density, $U$ is related to the thermodynamic property of the nematic LC and $L$ is related to the one-elastic-constant $K$. $W$ denotes the polar anchoring strength for the degenerate tangential anchoring on the boomerang surface. $\tilde{Q}_{ij} = Q_{ij} + \frac{1}{3}q_0\delta_{ij}$, where $q_0$ is the order parameter of the undistorted bulk LC and $\delta_{ij}$ is the Kronecker delta. $q_0$ and U are related by the following condition:

$$q_0 = \frac{1}{4} + \frac{3}{4}\sqrt{1 - \frac{8}{3U}}.$$

$\tilde{Q}_{ij}^{\perp} = P_{ik}\tilde{Q}_{kl}P_{lj}$, where $P_{\alpha\beta} = \delta_{\alpha\beta} - \nu_\alpha\nu_\beta$ is the projection operator associated with the surface normal $\nu$. In Eq.(8), the first term is the short-range Landau-de Gennes energy capturing nematic-isotropic phase transition, the second term is the elastic energy in one-constant approximation, and the last term is the Fournier-Galatola form for the degenerate tangential anchoring (*35*). To obtain a splay-bend director field in a rectangular channel (size $l_x \times l \times l_z$), we impose infinite anchoring walls at the $\pm x$ and $\pm z$ directions to force the director field to adopt the same pattern as in the experiments. A periodic boundary condition is applied in the *y*-direction. A boomerang with arm length 2.1 $\mu m$, arm width 0.55 $\mu m$, and thickness 0.51 $\mu m$, is then placed in the channel. A Ginzburg-Landau algorithm is performed to minimize the free energy (*36*). The numerical parameters are set as follows to match the experiments: $A = 1.17 \times 10^5$ J/m$^3$, $U = 3.0$, $L = 6$ pN, $W = 10^{-5}$ J/m$^2$, and pattern periodicity $l = 80$ μm.

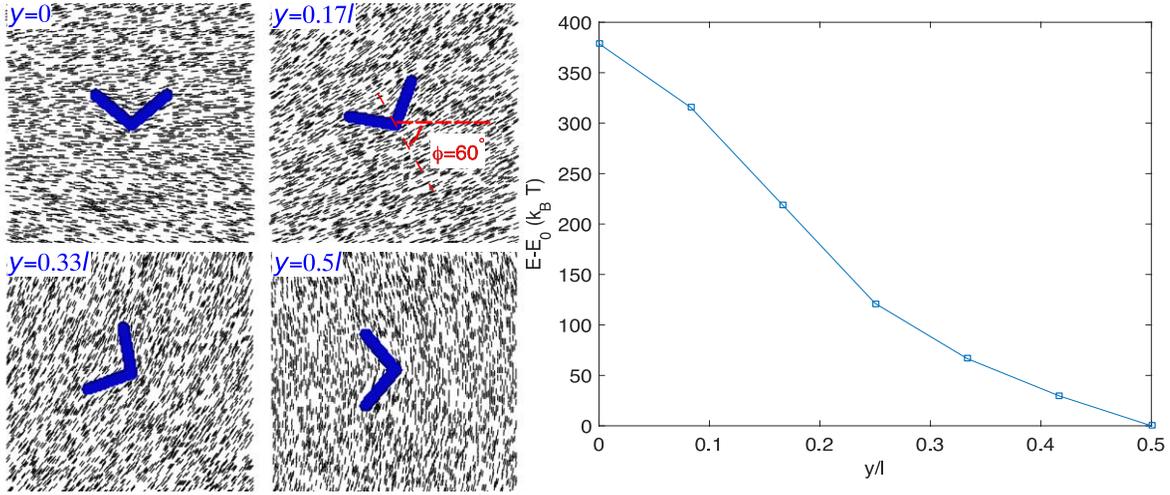

**Fig. 4. Simulation of boomerang particle in a photo-patterned cell.** On left, the director field is shown for four different positions. On right, the total free energy is plotted against the boomerang position. The orientation of the boomerang, characterized by the angle $\varphi$, is chosen to match the experimental observations. $\varphi = 90°, 60°, 30°, 0°$ for $y = 0, 0.17l, 0.33l, 0.5l$, respectively.

The simulation results are shown in Fig. 4. Because of the finite anchoring, there is only slight director alignment on the boomerang surface. We find that the system reaches the free energy minimum when $y = 0.5l$ with the boomerang orientation $\varphi = 0°$, consistent with the experimental observations. The free energy difference between $y = 0, \varphi = 90°$ and $y = 0.5l, \varphi = 0°$ is $\Delta E = 355 k_B T$, and the stiffness is $k = 4 \times 10^{-9}$ N/m. The free energy difference is about 2 times larger than the value obtained in the experiments. The prime reason is apparently the relatively high polar anchoring coefficient $W = 10^{-5}$ J/m$^2$ used in the simulations; as known from other studies, $W$ might be on the order of $10^{-6}$ J/m$^2$ (37).

**Sorting of colloids with different shapes.** As clear from the results presented above, the boomerang particles accumulate in the regions of maximum bend and align themselves with the symmetry plane being parallel to the *x*-axis. The placement and orientation of the boomerang is a direct result of its shape and the geometry of the surrounding director. Note that the angle between the two arms of the boomerang is larger than 90 degrees; for smaller angles, one might expect a different placement, in the splay regions.

It is of interest to compare the behavior of an obtuse-angle boomerang to that of spheres with perpendicular surface anchoring that also produce dipolar director distortions around themselves. We used polystyrene spheres with the radius $R = 2.5\,\mu\text{m}$ covered with a thin layer of a surfactant that yields perpendicular surface alignment of the director. When such a sphere is placed in a uniform nematic, the local radial-like director around the sphere is topologically compensated by a point defect, the so-called hyperbolic hedgehog located either on the right or left hand side of the sphere. Individual spheres with normal anchoring placed in the cell with a periodic splay and bend migrate towards the regions with the maximum splay, $y = 0, \pm l, \pm 2l, \ldots$, as has been already demonstrated in Ref. (29). When the cell is filled with a mixture of spherical and boomerang-shaped particles, the surface-patterned director landscape compartmentalizes them into different regions, despite the fact that in both cases the local distortions around the colloids are of a dipolar type, Fig. 5a. Namely, the normally anchored spheres migrate towards the maximum splay, $y = 0$, and the boomerang colloids migrate towards the maximum bend, $y = \pm l/2$, Fig. 5.

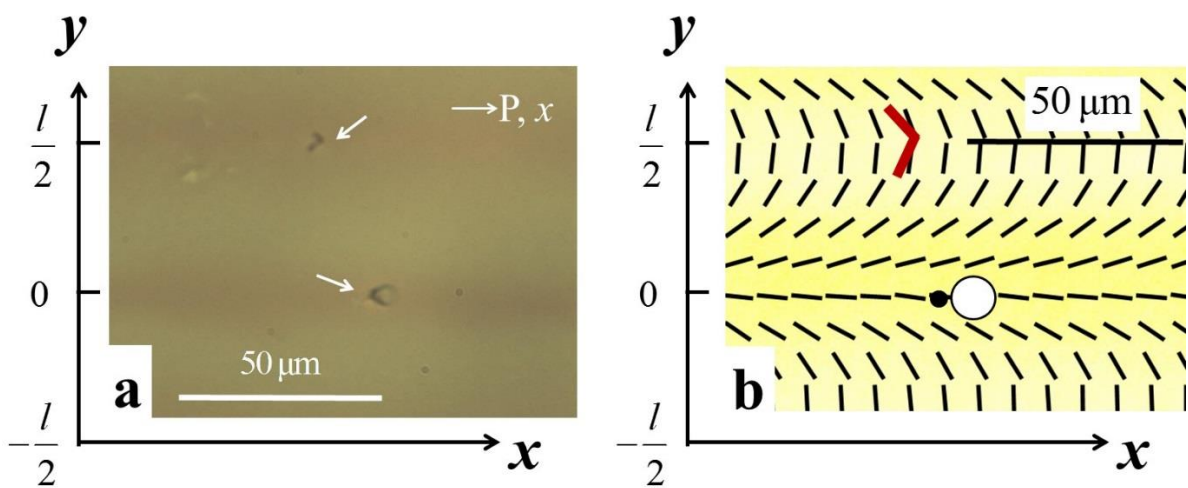

**Fig. 5. Preferential placement of spheres with normal anchoring and boomerangs with tangential anchoring in a LC cell with periodically patterned splay-bend director field.** (**a**) Microscope texture of polystyrene spherical colloids with homeotropic anchoring and boomerang colloid located in the alternating bands of splay and bend, respectively. (**b**) Scheme of preferential placement with the experimentally determined director pattern (PolScope observations). The normally anchored spheres reside in the region of splay, $y = 0$, while boomerang colloids localize in the regions of bend, $y = \pm l/2$. The colloids are placed in a LC mixture with zero dielectric anisotropy.

**Discussions and conclusions**

In this work, we demonstrate the effect of elastic interactions of nonspherical particles with the surrounding weakly distorted director pattern in the form of periodic splay and bend. The interactions lead to preferential placement and alignment of the boomerang particles. Namely, the particles migrate to the regions of a maximum bend and align in a polar fashion, with the curvature of the body adjusting to the curvature of the surrounding director. We quantify the strength of these interactions by measuring the empirical coefficient in the interaction potential. The interactions allow one to spatially separate colloids of different shapes, even if the director distortions created by them are of the same symmetry (dipolar in our case).

The separation is robust in the sense that the trapping potentials that keep the boomerang and spherical particles at different locations are very high, being two orders of magnitude higher than the thermal energy, see also Ref. (*29*). In this work, we used only the simplest one-dimensional director patterns in order to illustrate the concept. Because the system is invariant along the *x*-axis, the demonstrated preferential placement is restricted to the placement along the *y*-axis, i.e., the direction of the director modulations. To eliminate degeneracy along the $x$-axis placement, one needs to use two-dimensional director patterns, such as the ones described previously (*27, 28*).

Generally, the location and strength of the traps created by the patterned nematic landscape depends on three sets of variables: (i) the size, shape and surface anchoring details of the particle; (ii) the director pattern and surface anchoring strength of the bounding substrates and, finally, (iii) elastic parameters of the liquid crystal. In some cases, particles of different shape and surface anchoring type might find themselves trapped in similar locations. For example, one might expect that tangentially anchored spheres (*29*) and boomerang-shaped particles described in this work would both tend to locate in the regions of bend. Such a degeneracy can be mitigated or avoided if the sets (ii) and (iii) of parameters specified above are pre-designed taking into account the specific properties (i) of the colloids that need to be separated. In particular, the above-mentioned patterns with director gradients along both $x$ and $y$ axes are expected to expand the spectrum of colloids that can be placed in the desired pre-determined locations.

To conclude, the photo-patterning surface alignment offers a versatile method to control colloidal assemblies by directing particles of different shapes into predesigned locations and trapping them in these locations with forces that exceed those of entropic nature. Colloidal placement assembly in the photo-patterned templates can be realized at practically any spatial scale without mechanical stimuli (e.g., shearing, pressure gradients) or topographically modified surfaces (*38*). The described ability to deterministically pre-design locations of colloidal particles of different shapes in self-assembled patterns can be of importance for the development of microfluidic, electro-optical and sensing devices.

## Materials and Methods

We used the nematic MLC6815 and a LC mixture with zero dielectric anisotropy ($|\Delta\varepsilon| \leq 10^{-3}$) formulated by MLC7026-000 and E7 (in weight proportion 89.1:10.9). All the LCs are from *EM Industries*. The LCs were doped with 0.01wt% of boomerang colloids and polystyrene spheres (*Duke Scientific*). The spherical colloids treated with octadecyl-dimethyl-(3-trimethoxysilylpropyl) amsmonium chloride (DMOAP) produce perpendicular director alignment and dipolar director structures (*39*). The colloidal dispersion in the LC is injected into the photo-patterned cell with thickness $h = 20$ μm at room temperature 22°C.

## References

1. P. Poulin, H. Stark, T. C. Lubensky, D. A. Weitz, Novel colloidal interactions in anisotropic fluids. *Science* **275**, 1770-1773 (1997).
2. U. Tkalec, M. Ravnik, S. Čopar, S. Žumer, I. Muševič, Reconfigurable Knots and Links in Chiral Nematic Colloids. *Science* **333**, 62-65 (2011).
3. I. Muševič, M. Škarabot, U. Tkalec, M. Ravnik, S. Žumer, Two-Dimensional Nematic Colloidal Crystals Self-Assembled by Topological Defects. *Science* **313**, 954-958 (2006).
4. O. D. Lavrentovich, I. Lazo, O. P. Pishnyak, Nonlinear electrophoresis of dielectric and metal spheres in a nematic liquid crystal. *Nature* **467**, 947-950 (2010).
5. C. Blanc, D. Coursault, E. Lacaze, Ordering nano- and microparticles assemblies with liquid crystals. *Liquid Crystals Reviews* **1**, 83-109 (2013).
6. A. Sengupta, S. Herminghaus, C. Bahr, Liquid crystal microfluidics: surface, elastic and viscous interactions at microscales. *Liquid Crystals Reviews* **2**, 73-110 (2014).
7. G. Foffano, J. S. Lintuvuori, A. Tiribocchi, D. Marenduzzo, The dynamics of colloidal intrusions in liquid crystals: a simulation perspective. *Liquid Crystals Reviews* **2**, 1-27 (2014).
8. O. P. Pishnyak, S. Tang, J. R. Kelly, S. V. Shiyanovskii, O. D. Lavrentovich, Levitation, Lift, and Bidirectional Motion of Colloidal Particles in an Electrically Driven Nematic Liquid Crystal. *Physical Review Letters* **99**, 127802 (2007).
9. Y. Luo, F. Serra, K. J. Stebe, Experimental realization of the "lock-and-key" mechanism in liquid crystals. *Soft Matter* **12**, 6027-6032 (2016).
10. J. S. Lintuvuori, A. C. Pawsey, K. Stratford, M. E. Cates, P. S. Clegg, D. Marenduzzo, Colloidal Templating at a Cholesteric-Oil Interface: Assembly Guided by an Array of Disclination Lines. *Physical Review Letters* **110**, 187801 (2013).
11. D. Voloschenko, O. P. Pishnyak, S. V. Shiyanovskii, O. D. Lavrentovich, Effect of director distortions on morphologies of phase separation in liquid crystals. *Physical Review E* **65**, 060701 (2002).
12. D. K. Yoon, M. Choi, Y. H. Kim, M. W. Kim, O. D. Lavrentovich, H. T. Jung, Internal structure visualization and lithographic use of periodic toroidal holes in liquid crystals. *Nature Materials* **6**, 866-870 (2007).
13. D. Pires, J.-B. Fleury, Y. Galerne, Colloid Particles in the Interaction Field of a Disclination Line in a Nematic Phase. *Physical Review Letters* **98**, 247801 (2007).
14. H. Yoshida, K. Asakura, J. Fukuda, M. Ozaki, Three-dimensional positioning and control of colloidal objects utilizing engineered liquid crystalline defect networks. *Nature Communications* **6**, 7180 (2015).

15. X. Wang, D. S. Miller, E. Bukusoglu, J. J. de Pablo, N. L. Abbott, Topological defects in liquid crystals as templates for molecular self-assembly. *Nat Mater* **15**, 106 (2016).
16. M. Mitov, C. Portet, C. Bourgerette, E. Snoeck, M. Verelst, Long-range structuring of nanoparticles by mimicry of a cholesteric liquid crystal. *Nat Mater* **1**, 229-231 (2002).
17. A. Nych, U. Ognysta, M. Škarabot, M. Ravnik, S. Žumer, I. Muševič, Assembly and control of 3D nematic dipolar colloidal crystals. *Nat Commun* **4**, 1489 (2013).
18. T. A. Wood, J. S. Lintuvuori, A. B. Schofield, D. Marenduzzo, W. C. K. Poon, A Self-Quenched Defect Glass in a Colloid-Nematic Liquid Crystal Composite. *Science* **334**, 79-83 (2011).
19. B. I. Senyuk, Q. Liu, S. He, R. D. Kamien, R. B. Kusner, T. C. Lubensky, I. I. Smalyukh, Topological colloids. *Nature* **493**, 200-205 (2013).
20. M. Zapotocky, L. Ramos, P. Poulin, T. C. Lubensky, D. A. Weitz, Particle-Stabilized Defect Gel in Cholesteric Liquid Crystals. *Science* **283**, 209 (1999).
21. K. Higashiguchi, K. Yasui, M. Ozawa, K. Odoi, H. Kikuchi, Spatial distribution control of polymer nanoparticles by liquid crystal disclinations. *Polym J* **44**, 632-638 (2012).
22. X. Wang, Y.-K. Kim, E. Bukusoglu, B. Zhang, D. S. Miller, N. L. Abbott, Experimental Insights into the Nanostructure of the Cores of Topological Defects in Liquid Crystals. *Physical Review Letters* **116**, 147801 (2016).
23. D. Coursault, J. Grand, B. Zappone, H. Ayeb, G. Lévi, N. Félidj, E. Lacaze, Linear Self-Assembly of Nanoparticles Within Liquid Crystal Defect Arrays. *Advanced Materials* **24**, 1461-1465 (2012).
24. D. Kasyanyuk, P. Pagliusi, A. Mazzulla, V. Reshetnyak, Y. Reznikov, C. Provenzano, M. Giocondo, M. Vasnetsov, O. Yaroshchuk, G. Cipparrone, Light manipulation of nanoparticles in arrays of topological defects. *Scientific Reports* **6**, 20742 (2016).
25. J.-B. Fleury, D. Pires, Y. Galerne, Self-Connected 3D Architecture of Microwires. *Physical Review Letters* **103**, 267801 (2009).
26. V. M. Pergamenshchik, Elastic multipoles in the field of the nematic director distortions. *The European Physical Journal E* **37**, 1-15 (2014).
27. C. Peng, Y. Guo, C. Conklin, J. Viñals, S. V. Shiyanovskii, Q.-H. Wei, O. D. Lavrentovich, Liquid crystals with patterned molecular orientation as an electrolytic active medium. *Physical Review E* **92**, 052502 (2015).
28. Y. Guo, M. Jiang, C. Peng, K. Sun, O. Yaroshchuk, O. Lavrentovich, Q.-H. Wei, High-Resolution and High-Throughput Plasmonic Photopatterning of Complex Molecular Orientations in Liquid Crystals. *Advanced Materials* **28**, 2353–2358 (2016).
29. C. Peng, T. Turiv, Y. Guo, S. V. Shiyanovskii, Q.-H. Wei, O. D. Lavrentovich, Control of colloidal placement by modulated molecular orientation in nematic cells. *Science Advances* **2**, e1600932 (2016).
30. S. C. Glotzer, M. J. Solomon, Anisotropy of building blocks and their assembly into complex structures. *Nat Mater* **6**, 557-562 (2007).
31. C. P. Lapointe, T. G. Mason, I. I. Smalyukh, Shape-Controlled Colloidal Interactions in Nematic Liquid Crystals. *Science* **326**, 1083-1086 (2009).
32. B. Senyuk, Q. Liu, E. Bililign, P. D. Nystrom, I. I. Smalyukh, Geometry-guided colloidal interactions and self-tiling of elastic dipoles formed by truncated pyramid particles in liquid crystals. *Physical Review E* **91**, 040501 (2015).
33. A. Chakrabarty, A. Konya, F. Wang, J. V. Selinger, K. Sun, Q.-H. Wei, Brownian Motion of Boomerang Colloidal Particles. *Physical Review Letters* **111**, 160603 (2013).
34. P. De Gennes, J. Prost, *The physics of liquid crystals, 2nd ed.*, (Clarendon, Oxford, 1993).
35. J. B. Fournier, P. Galatola, Modeling planar degenerate wetting and anchoring in nematic liquid crystals. *EPL (Europhysics Letters)* **72**, 403 (2005).
36. M. Ravnik, S. Žumer, Landau–de Gennes modelling of nematic liquid crystal colloids. *Liquid Crystals* **36**, 1201-1214 (2009).
37. D. Andrienko, Y. Kurioz, Y. Reznikov, C. Rosenblatt, R. Petschek, O. Lavrentovich, D. Subacius, Tilted photoalignment of a nematic liquid crystal induced by a magnetic field. *Journal of Applied Physics* **83**, 50-55 (1998).
38. N. M. Silvestre, Q. Liu, B. Senyuk, I. I. Smalyukh, M. Tasinkevych, Towards Template-Assisted Assembly of Nematic Colloids. *Physical Review Letters* **112**, 225501 (2014).
39. I. Lazo, C. Peng, J. Xiang, S. V. Shiyanovskii, O. D. Lavrentovich, Liquid crystal-enabled electro-osmosis through spatial charge separation in distorted regions as a novel mechanism of electrokinetics. *Nature Communications* **5**, 5033 (2014).

**Acknowledgments**

**Funding:** The experiments presented in this work were supported by the National Science Foundation grant DMR-1507637. The simulations and theoretical analysis presented in this work were supported by the National Science Foundation through grant DMR-1121288. Preparation of plasmonic photomasks was supported by the National Science Foundation through grant CMMI-1436565. **Competing interests:** The authors declare no competing interests. **Data and materials availability:** All data needed to evaluate the conclusions in the paper are present in the paper. Additional data related to this paper may be requested from the authors.